# GPS Receiver with Enhanced User Positioning Time

Seung-Hyun Yoon[*], Ji-Woon Jung, Su-Bong Kim, Hyun-Chang Shin, Jae-Hyang Lee, Kyu-Yun Lee, and Hyo-Sun Shim

*Digital Media R&D Center, Samsung Electronics*

**ABSTRACT**

This paper introduces a Global Positioning System (GPS) Receiver that locates user's position instantly. Recently, many mobile devices require location information to add user position into their contents, and some applications require quick positioning when the device is initially switched on.

In order to reduce the time to fix user's position, we propose the Instant-On GPS receiver system which is implemented on an ARM based FPGA board, and operates under a very low power mode. We've developed a repeated sleep mode by periods to control the GPS receiver's main power in order to achieve reduced power consumption. By using a high resolution Real Time Clock (RTC), we can estimate frame sync timing without receiving the current frame sync preamble data from a satellite when GPS turns back on. However, the navigation solution needs to be calculated once in advance.

The performance results of the proposed GPS receiver in both real world and simulation environment are presented.

## 1 INTRODUCTION

The new development in Location Based Service (LBS) business will lead personal mobile entertainment devices to be equipped with positioning solutions of their own, such as GPS or network assisted positioning. One major concern that these systems have is when the systems are needed to be operational abruptly. A personal mobile device normally requires a waiting time of around 30 s to acquire user's position information that can be added onto its contents using the conventional GPS system. This waiting time limits the usability of GPS in personal mobile devices such as Digital Still Camera (DSC) and Cellular phone that are used briefly at a time. Launching Assist-GPS (AGPS) is one of possible techniques to reduce time to locate user's position. However, it requires additional global reference network for receiving satellite information such as ephemeris, satellite health, status and almanac data. It also needs a server infrastructure to gather satellite information and forward it to GPS receiver. To run AGPS, personal mobile devices have to be equipped with a network interface to communicate with the server infrastructure. These additional requirements make each personal mobile device to become larger in size and to consume more power.

The objective of this paper is to introduce a GPS receiver with reduced user positioning time. The GPS receiver is managed to remain at ready state to achieve navigation solution that locates its position instantly; therefore, the user can find out about his/her location without waiting for 30 s. Section 2 is an overview of the GPS. Section 3 presents a technique to reduce the time to locate user's position using a RTC and a frame sync estimator. Section 4 shows the simulation result and real environment test result of this technique. Finally, Section 5 is with conclusions.

## 2 GPS SIGNAL STRUCTURE

GPS satellites transmit precisely timed GPS signals in two $L$-band frequencies 1575.42 MHz and 1227.6 MHz. These signals must have embedded in them, in the form of navigation data, both the precise satellite clock time as well as satellite position so that a user receiver can determine both satellite time and satellite position at the time of transmission. The $L_1$ signal is Binary Phase Shift Key (BPSK) modulated with Pseudo Random Noise (PRN) code plus 50 bits per second navigation data. Codes for various satellites are taken from a family of codes known as Gold codes and the code has a chip rate of 1.023 MHz. Each code is unique and provides the mechanism to identify each satellite in the constellation.

The 50 bits per second rate navigation data transmitted by the satellites are formatted into 30-bit words, and the words are grouped into sub-frames of 10 words that are 300 bits in length and 6 s in duration. Frames consist of 5 sub-frames of 1500 bits and 30 s in duration, and a super-frame consists of 25 frames and has duration of 12.5 min.

Each satellite transmits its independent information such as position, parity check data, ephemeris data through sub-frame 1st to 3rd. Sub-frame 4th and 5th is almanac data for all satellites information in the constellation [1, 3].

If the receiver's clock and satellites' clocks are synchronized and more than four satellite signals are tracked, then precise user position can be calculated. In order to track satellite signals, Doppler frequency and PRN code phase must be found. Each satellite has a different Doppler frequency related with its current position, current user position and velocity. It also has a different PRN code and code phase rotates with period.

Once the correlation power between satellite signals and receiver generated PRN codes which include Doppler frequency reduction process is above the threshold value, acquisition engine declares the code lock. If both the Doppler frequency reduction process and the PRN code phase matching process are in success and correlation power keeps in state above the threshold value, carrier lock is declared. Bit lock is declared when 50bps bit transition can be achieved. After frame sync preamble is decoded from navigation data, at last, frame lock is declared.

As mentioned before, the 1st to the 3rd sub-frame data of more than 4 satellites are needed to calculate precise user position. It takes 18 s to receive mathematically because each frame consists of 6 s data set. However each satellite has different position and related velocity from the receiver which generates difference in propagation delays between each satellite and the receiver. Therefore the receiver can't receive sub-frame preambles of four satellites at the same time even though each satellite transmits it simultaneous. Normally it takes 18 s to 30 s for a user's receiver to receive the 1st to the 3rd sub-frame data from four satellites. If AGPS technique is applied, ephemeris data is acquired from another network source. Therefore, the user's position can be calculated right after the receiver declares frame lock.

## 2.1 PSEUDORANGE CALCULATION

Calculating precise user position needs more than four satellites' positions which are known from the ephemeris data transmitted by the satellites, and pseudo range from user receiver to the satellites. With this information, user's position can be determined through following equations.

$$\begin{aligned} \rho_1 &= \sqrt{(x_1 - x_u)^2 + (y_1 - y_u)^2 + (z_1 - z_u)^2} + b_u \\ \rho_2 &= \sqrt{(x_2 - x_u)^2 + (y_2 - y_u)^2 + (z_2 - z_u)^2} + b_u \\ \rho_3 &= \sqrt{(x_3 - x_u)^2 + (y_3 - y_u)^2 + (z_3 - z_u)^2} + b_u \\ \rho_4 &= \sqrt{(x_4 - x_u)^2 + (y_4 - y_u)^2 + (z_4 - z_u)^2} + b \end{aligned} \quad (1)$$

where $x_n, y_n, z_n$ are user receiver position and $b_u$ is the user clock bias error expressed in distance. Four equations are needed to solve four unknowns: $x_u, y_u, z_u$ and $b_u$. Satellite positions $x_n, y_n, z_n$, ($n$=1, 2, 3 and 4) are acquired from ephemeris data. Pseudoranges $\rho_i$, ($n$=1, 2, 3 and 4) are calculated through following sequences. Every satellite sends a frame sync preamble at a certain time $t_{si}$ which is synchronized to zero second in GPS time. The receiver will receive the signal at a later time $t_u$. The distance between the user and the satellite $i$ is

$$\rho_i = c(t_u - t_{si}) \quad (2)$$

where $c$ is the speed of light, $\rho_i$ is often referred to as the true value of pseudorange from user to satellite $i$, $t_{si}$ is referred to as the true time of transmission from satellite $i$, $t_u$ is the true time of reception [2]. The receiver declares frame sync when it receives frame preambles which start satellites at certain time zero second. The receiver has a receiver clock and a TIC which occurs at 100 ms interval. When the receiver gets frame data, precise counter starts to count code chips passing through from the first receiving time to the receiver TIC occurring time. The receiver stores this value as a code time then converts the receiver time to GPS time. Propagation delay can be calculated with information such as frame data start time, code time and GPS time at the TIC. Finally pseudorange is calculated by multiplying propagation delay time with the speed of light constant.

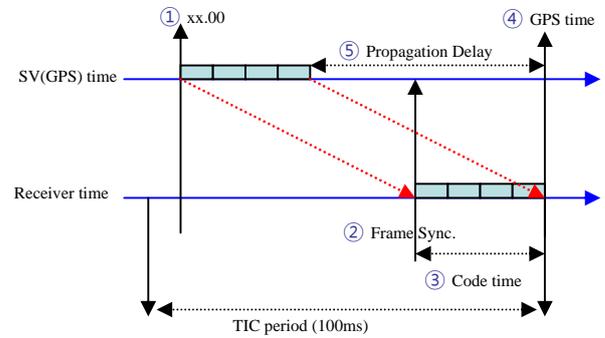

Fig. 1 Pseudorange calculation.

## 2.2 CONVERT RECEIVER TIME TO GPS TIME

Converting receiver time to GPS time at the TIC takes following steps. First, calculate Receiver Clock Offset (RCO) value which indicates the time offset between receiver time and GPS time. The GPS time can be acquired by subtracting RCO value from current receiver time.

RCO is updated regularly when the TIC occurs right after receiver acquires frame sync preambles. It can be expressed as

RCO.week = ZT.week - WeekNumber
RCO.seccond = ZT.seccond + SyncTIC × 0.1 – (TOW×6
   + 0.075)                                                   (3)

where ZT is the zero time of the receiver which indicates the time when the receiver is initially started, WeekNumber is current week number synchronized to GPS time obtained from satellite navigation message, SyncTIC is expected TIC value at the end of sub-frame currently being received, TOW indicates the GPS time at the end of current sub-frame obtained from satellite navigation message. When TIC occurs after the receiver gets frame sync preambles, the receiver can calculate the TIC value at the end of current sub-frame because each sub-frame is consist of exactly 10 words and every first word has a frame sync preamble. The second word of each sub-frame has a TOW data which is updated by satellites in 6 s resolution and indicates the GPS time at the end of current sub-frame. Comparing these two values, where both indicate the time at the end of current sub-frame, RCO value can be calculated.

## 3 FRAME SYNC ESTIMATOR

As mentioned before, frame sync preambles, TOW and WeekNumber values should be received from satellites to update the RCO value in the receiver. With updated RCO value and ephemeris data from sub-frame, user position can be calculated. Each ephemeris data has a time validation period, so it can be reused within 4 hours after receiving once without updating full ephemeris data from satellites which takes at lease 18 s for sub-frames 1st to 3rd. The first word of each sub-frame has frame sync preamble and the second word has TOW value. When the receiver is turned on, it starts to receive satellite data from certain points between 1st word and 10th word. It will take 1.2 s to 6 s depends on the case to receive frame sync preambles and TOW values. So, it will take the same amount of time that is 1.2 s to 6 s to acquirer user's position even though the receiver has a validate ephemeris data.

With adding precise counter using a Real Time Clock (RTC) and a frame sync estimator block to the receiver, TOW and SyncTIC can be estimated right after bit lock declaration. GPS time at the end of current sub-frame also can be calculated without receiving frame sync preambles by estimating current bit and word index using RTC counter differences during turning off process of the receiver. It will reduce the time to fix user's position to near zero second except for the time spent on code, carrier and bit lock. To operate frame sync estimator properly, current word index, bit index, TOW value and RTC counter value should be stored before the receiver goes to turn off state. When the receiver is turned back on by a user action, current word index, bit index and TOW value are estimated based on increment of RTC during the turning off state after carrier, code and bit lock are declared. With this estimated information, frame sync is declared instantly without waiting for receiving frame sync preambles, and user's position is calculated simultaneously. Fig. 2 shows the overall flow of calculating user's position.

For example, current word counter is 6, bit counter is 19, TOW is 2679 and RTC counter is 17362. The RTC counter uses 32 kHz clock to add up counts. The receiver turns off with this information stored and turns back on later. After the time code, carrier and bit lock are declared, the RTC counter value is 673176. With the discrepancy in RTC value, the turning off state time can be calculated by (6731176 – 17362) / 32 = 209806.6875 ms. This value is categorized in three parts as 600ms $\times$ 349 + 20 ms $\times$ 20 + 1 ms $\times$ 6.6875. The information for calculating user's position are estimated as follows

estimated bit index = (19 + 20) modulus 30 = 9
estimated word index = (6 + 349 + 1) modulus 10 = 6
estimated TOW = 2679 + INT(209806.6875 / 6000) = 2713
estimated SyncTIC = current TIC + (10 – Current word counter) $\times$ 6        (4)

There are two important factors to be considered in using frame sync estimator block. First, use high quality RTC. Frame sync estimator block operates based on the difference between RTC counters. RTC frequency tolerance is important parameter to decide the period that the receiver can be turned off. If the RTC frequency tolerance is 10 ppm and bit index margin is 10 ms, turning off time can expressed

$$10 / 32 \text{ kHz (ms)} > \text{counts} / 32 \text{ kHz} \times 10 / 10^6 \text{ (ms)}$$
$$\text{counts} < 10^6 \text{ ms} \qquad (5)$$

With this RTC frequency tolerance, the receiver can be turned off for $10^6$ ms. During turning off state, mobile devices can save its power. In the case where the receiver remains at the turning off state for 15 minutes and re-operates for 2 s to 3 s, the receiver will achieve low power consumption that consumes a 300th of the power consumed by the conventional GPS receiver. Another factor engaged in the performance of frame sync estimator block is the code Doppler frequency. Normally the GPS signals add the carrier Doppler frequency $\pm 10$ kHz. The GPS carrier frequency that the receiver use is 1575.42 MHz and PRN code rate is 1.023 MHz. So each PRN code can have $\pm 6.49$ Hz code Doppler frequency. When the code and the carrier lock are declared before frame sync estimator runs, carrier Doppler frequency can be acquired. This current Doppler frequency and the past Doppler frequency, that is stored when the receiver goes to the turning off state, are provided to frame sync estimator block to compensate code Doppler effects.

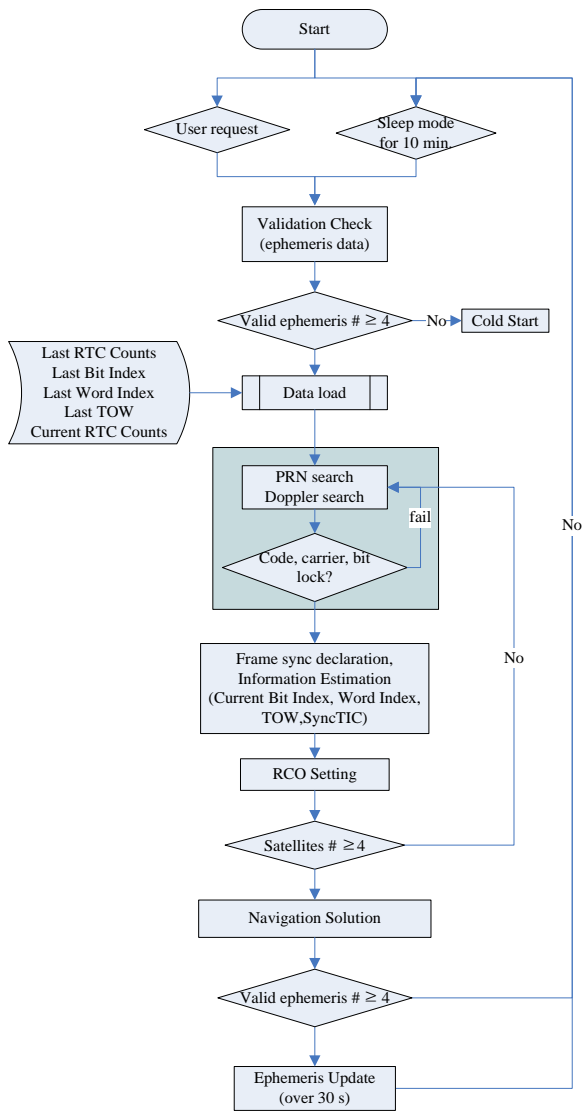

Fig. 2 Flow chart of locating user's position

## 4 PERFORMANCE EVALUATION

The frame sync estimator is used to locate user's position instantly after carrier, code and bit lock is declared. Since the main objective of this research is to develop and test a new algorithm, the data is recorded only when the receiver can acquire enough satellites with a high signal power. Fig. 3 shows the data collection setup. Both signals received from a signal generator and real environment are used for testing.

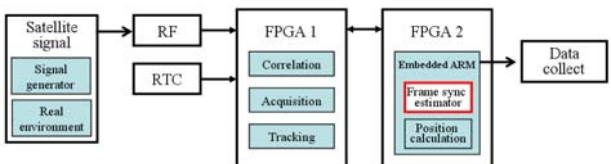

Fig. 3 Test environment block diagram

For evaluating the frame sync estimator in simulation environment, the equipments resented in Table 1 are used.

Table 1. Test set-up for evaluating the frame sync estimator

| Equipments | Details |
|---|---|
| Signal Generation | GSS6560 (Spirent) |
| RF Front-end | SE4120 (SiGe) |
| Receiver | Proposed GPS receiver L1 C/A |
| RTC | ACLOL (ABRACON) |
| FGPA1 | HAPS-54 (Synplicity) Virtex-5 LX330 |
| FPGA2 | M1A3PE1500 (Actel) |
| Tracked Satellite number before turning off | 8 |
| Satellite signal power | -130 dBm |

Fig. 4 depicts the frame sync estimator performance result in terms of time delay to locate user's position after turning the receiver back on. The frame sync estimator algorithm is compared to the conventional Hot Start method. When the receiver wakes up, it sets initial user position with the fixed value of the last known position which is stored before it is turned off. In the figure, the first data after zero, 2-Dimensional Root Mean Square (2D-RMS) pseudorange error presents the error value with the last fixed position data. Then the receiver starts to calculate user's position. It takes approximately 1 s for the frame sync estimator to find the solution. The 2D-RMS pseudorange error indicates the accuracy of proposed GPS receiver, and the x-axis has an 1 s time step.

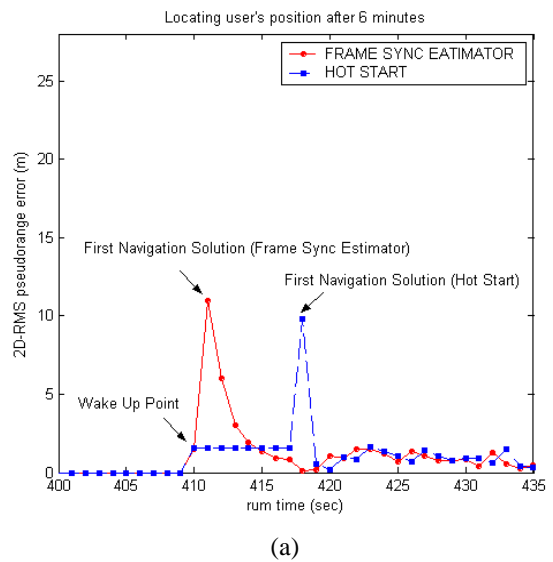

(a)

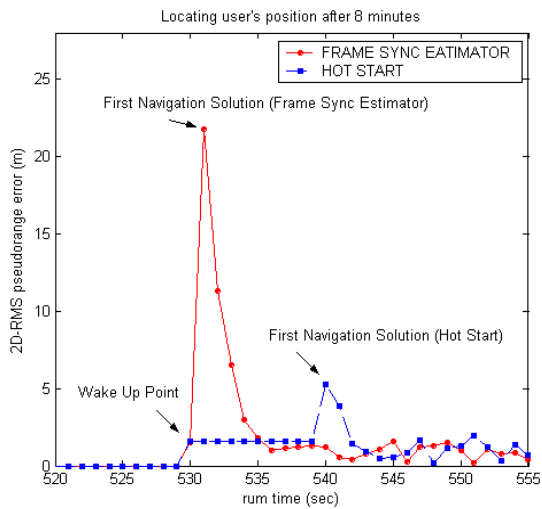

(b)

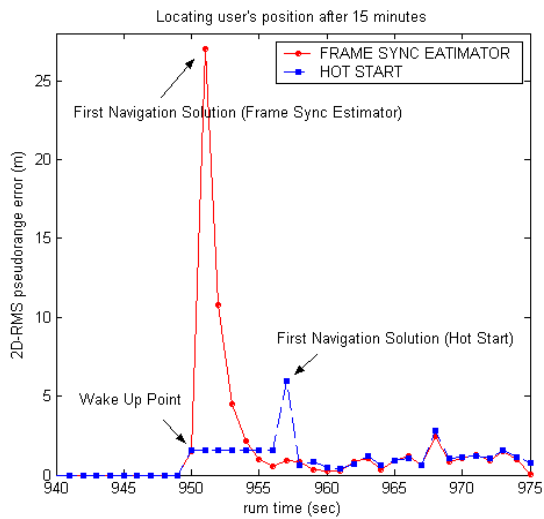

(c)

Fig. 4 Locating user's position using Frame sync estimator and Hot Start; (a) turning on after 6 minutes, (b) turning on after 8 minutes, (c) turning on after 15 minutes

### 4.1 FIELD TEST

The proposed frame sync estimator was tested in real environment. First the GPS platform was turned on and waited to locate user's position. The test was progressed when it has 8 satellites to ensure satellite signal quality. Then it was turned off for 15 minutes. At last, it was turned back on, and navigation data was recorded into a file. The Hot Start method was also tested to be compared with. Fig. 5 shows the result. The GPS receiver with frame sync estimator can locate user's position within 1 s. The first navigation solution has a large 2D-RMS pseudorange error due to the fact that the receiver assumed the propagation delay time from satellite of 0.075 s at the first time. Then it used more accurate propagation delay time from the previous navigation solution. The 2D-RMS pseudorange error decreased to below 10 m within 2 s.

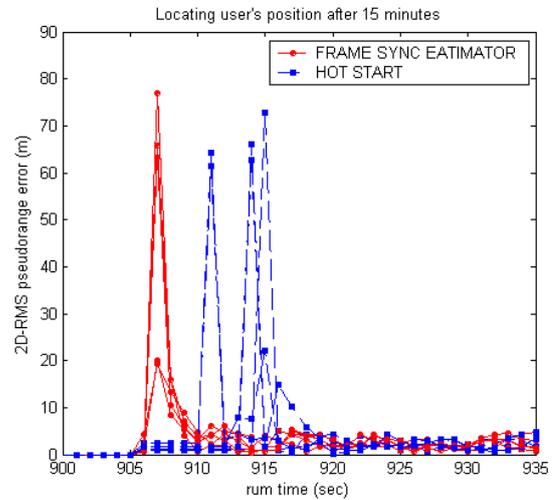

Fig. 5 Time delay to locate user's position using frame sync estimator and Hot Start method in real environment.

### 5 CONCLUSIONS

This paper has discussed the implementation of a new type of GPS receiver with enhanced user positioning time. Using frame sync estimator, the receiver doesn't need to wait for frame sync preambles. So it can locate user's position within 1 s with 20 to 80 m 2D-RMS pseudorange error value. After 1 s from the first navigation solution, it calculates user position with high accuracy that is within 10 m 2D-RMS pseudorange error. The performance of the frame sync estimator and the conventional Hot Start mechanism are compared in terms of the time delay and 2D-RMS accuracy, and the proposed GPS receiver has shown a better performance.